\def\year{2022}\relax
\documentclass[letterpaper]{article} 
\usepackage{aaai22} 
\usepackage{times} 
\usepackage{helvet} 
\usepackage{courier} 
\usepackage[hyphens]{url} 
\usepackage{graphicx} 
\usepackage{natbib} 
\usepackage{caption} 
\DeclareCaptionStyle{ruled}{labelfont=normalfont,labelsep=colon,strut=off} 
\usepackage{algorithm}
\usepackage{algorithmic}

\usepackage{makecell} 
\usepackage{amsmath}
\usepackage{newfloat}
\usepackage{listings}
\floatstyle{ruled}
\newfloat{listing}{tb}{lst}{}
\floatname{listing}{Listing}

\title{De-risking Carbon Capture and Sequestration with Explainable CO${_2}$ Leakage Detection in Time-lapse Seismic Monitoring Images}
\author{
  Huseyin Tuna Erdinc,\equalcontrib\textsuperscript{\rm 1}
  Abhinav Prakash Gahlot,\equalcontrib \textsuperscript{\rm 2}
  Ziyi Yin,\textsuperscript{\rm 3} \\
  Mathias Louboutin,\textsuperscript{\rm 2}
  Felix J. Herrmann\textsuperscript{\rm 1,2,3}
}
\affiliations{
  \textsuperscript{\rm 1}School of Electrical and Computer Engineering, Georgia Institute of Technology\\
  \textsuperscript{\rm 2}School of Earth and Atmospheric Sciences, Georgia Institute of Technology\\
  \textsuperscript{\rm 3}School of Computational Science and Engineering, Georgia Institute of Technology\\
  \{herdinc3, agahlot8, ziyi.yin, mlouboutin3, felix.herrmann\}@gatech.edu,

}

\usepackage{bibentry}

\begin{document}

\maketitle

\begin{abstract}

With the growing global deployment of carbon capture and sequestration technology to combat climate change, monitoring and detection of potential CO${_2}$ leakage through existing or storage induced faults are critical to the safe and long-term viability of the technology. Recent work on time-lapse seismic monitoring of CO${_2}$ storage has shown promising results in its ability to monitor the growth of the CO${_2}$ plume from surface recorded seismic data. However, due to the low sensitivity of seismic imaging to CO${_2}$ concentration, additional developments are required to efficiently interpret the seismic images for leakage. In this work, we introduce a binary classification of time-lapse seismic images to delineate CO${_2}$ plumes (leakage) using state-of-the-art deep learning models. Additionally, we localize the leakage region of CO${_2}$ plumes by leveraging Class Activation Mapping methods.

\end{abstract}

\section{Introduction}

According to the International Energy Agency and the International Panel on Climate Change report \citep{IPCC2018}, there is a  need for a 50 percent reduction of greenhouse gas emissions by 2050 to avoid an increase of 1.5 degrees Celsius of Earth’s average temperature. This can only be achieved by reduced dependence on fossil fuels, use of renewable sources of energy and large-scale global deployment of carbon reduction technologies such as carbon capture and sequestration (CCS). This technology consists of collection, transportation, and injection of CO${_2}$ into an appropriate geologic storage reservoir for extended time periods (tens of years). Especially, unlike other solutions, CCS is considered a relatively low-cost, long-term and imminent solution. However, potential CO${_2}$ leakage from the underground reservoirs due to pre-existing or pressure-induced faults poses risks \cite{ringrose2020store}. Thus, it is necessary to de-risk CCS projects by monitoring CO${_2}$ plumes in order to accurately detect and predict potential leakages as early as possible.

Time-lapse seismic monitoring has been introduced as a reliable technology to monitor the CO${_2}$ dynamics in the Earth's subsurface during carbon sequestration \cite{doi:10.1190/1.1444921} and is already in use at existing storage sites \cite{Arts2008TenYE, doi:10.1190/1.3304820, RINGROSE20136226,FURRE20173916}. In essence, sequential (i.e once every 6 months/year/...) seismic datasets, called vintages, are collected in the field over an area covering the storage reservoir. Then, each seismic dataset is inverted to obtain high fidelity images of the subsurface over time \cite{Arts2008TenYE, Ayeni&Biondi, yin2021SEGcts}. The evolution of the CO${_2}$ reservoir can finally be visualized  by subtracting the seismic images between different points in time. However, due to the inherently weak and noisy amplitudes of the CO${_2}$ reservoir's response in those seismic images, detecting the presence of potential irregularities, such as in CO${_2}$ plumes, corresponding to a leakage is a challenging problem. To tackle this difficulty, we propose a machine learning based detection method based on standard binary classification.

Recently, numerous methods leveraging machine learning have been introduced for the detection of CO${_2}$ leakage based on a simple artificial neural network (ANN) \cite{LI2018276}, and a combination of convolutional neural networks (CNN) and Long Short-Term Memory (LSTM) networks \cite{ZHOU2019102790}. While leading to accurate predictions, these methods usually rely solely on the field recorded data rather than the subsurface seismic images. Besides, practical considerations such as repeatability (the ability to record the data in the exact same way every year) hinders their applicability to real world cases. On the other hand, as we rely on visualizing the CO${_2}$ plumes in the seismic image, we can take advantage of advanced seismic imaging techniques designed for non-repeated seismic acquisition such as the joint recovery model (JRM) \cite{oghenekohwo2017hrt, wason2016GEOPctl, yin2021SEGcts}. Additionally, this imaging technique has demonstrated higher fidelity imaging than sequential seismic imaging allowing for easier detection of CO${_2}$ leakage.

We will show in the following sections that we can efficiently and accurately detect CO${_2}$ from realistic seismic images recovered by JRM on synthetic but representative models of the Earth subsurface. We demonstrate our method using different state-of-the-art deep learning models in a transfer learning setting to classify CO${_2}$ plume seismic images with regular (no-leakage) CO${_2}$ plume or with CO${_2}$ leakage. As CO${_2}$ leakage detection needs trustworthiness, we further unravel the decisions made by our models and utilize Class Activation Mapping (CAM) methods \cite{cam} to identify and visualize seismic image areas crucial for model classification results. We show that the CAM result accurately focuses on the leakage portion of the CO${_2}$ plume and reservoir, validating that our network detects leakage based on state of the CO${_2}$ reservoir over time.

Our main contributions are organized as follows. First, we introduce the classification models used for leakage detection and the CAM methods for visualizing the area of interest in the seismic image. Second, we demonstrate the accuracy of our models and qualitatively examine the results of our CAM methods on a realistic synthetic set of CO${_2}$ plume images.

\section{Methodology}

In order to speed up the training process and to compensate for the overfitting that may occur with modest sized datasets, we rely on transfer learning \cite{transfer_learning} using pre-trained state-of-the-art models as a starting point. In particular, we consider four modern architectures known to achieve high accuracy on standard dataset such as ImageNet-1k \cite{imagenet}. The models used are VGG \cite{vgg}, ResNet \cite{resnet}, Vision Transformer (ViT) \cite{Visiontransformer}, and Swin Transformer (Swin) \cite{Swintransformer}, all pre-trained on the standardized ImageNet-1k dataset.

\textbf{VGG:} is a convolutional neural network (CNN) model that achieved significant success in The ImageNet Large Scale Visual Recognition Challenge (ILSVRC) competition in 2014 \cite{vgg}. VGG consists of sequences of convolution and maxpool layers. In our numerical experiments, the VGG16 variant with 16 trainable layers is used.

\textbf{ResNet:} is a CNN architecture with residual connections proposed to solve the vanishing gradient problem in very deep networks \cite{resnet}. ResNet consists of residual blocks and each residual block has convolution layers and shortcut connections performing identity mapping. In our numerical experiments, the ResNet34 variant with 34 trainable layers is used.

\textbf{ViT:} is an architecture based on transformer which is used in the field of Natural Language Processing (NLP) \cite{transformer}. Internally, the transformer learns a relationship between input token pairs, and uses 16x16 patches of images as input tokens \cite{Visiontransformer}. In our numerical experiments, the tiny ViT variant is used allowing lower memory and computational imprint.

\textbf{Swin:} is a special type of ViT that represents image patches hierarchically by starting from small-sized patches and gradually increasing the size through merging to achieve scale-invariance property \cite{Swintransformer}. Compared to ViT, Swin transformer has superior (linear) computational efficiency by computing self-attention within certain patches of a window. In our numerical experiments, tiny Swin variant is used allowing lower memory and computational imprint.

\subsection{CAM Methods}

Deep learning models for classification are notoriously treated as ``black boxes'' as they do not expose their internal knowledge or operations to its users and do not provide interpretable results. To solve this problem, CAM based saliency maps (heatmaps) were introduced to highlight the most class-discriminative regions of to-be-classified input images \cite{cam}. Since CO${_2}$ leakage requires high fidelity, transparent and interpretable models, we use CAM to further make our model results explainable and highlight the regions of the seismic image that are most relevant to the classification results. In our study, we considered two CAM methods. First, Grad-CAM \cite{Grad-CAM}, a gradient-based CAM method considered as the state-of-the-art in terms of explainability of neural networks for classification. This CAM method extracts gradients from a specific layer of a model and computes the weighted average of that specific layer's activations. Second, we consider Score-CAM \cite{ScoreCAM}, a perturbation based CAM method. Score-CAM also computes the weighted average of activations of a user-specified layer but, unlike Grad-CAM, Score-CAM relies on propagating (forward pass through the network) a masked input image where the mask is obtained via upsampling the activations of the user-defined layer. This CAM method provides high accuracy and interpretable heatmaps and alleviates potential noise and spread present in the gradient used for the Grad-CAM heatmaps.

\begin{figure*}[t]
\centering
\includegraphics[width=0.85\textwidth]{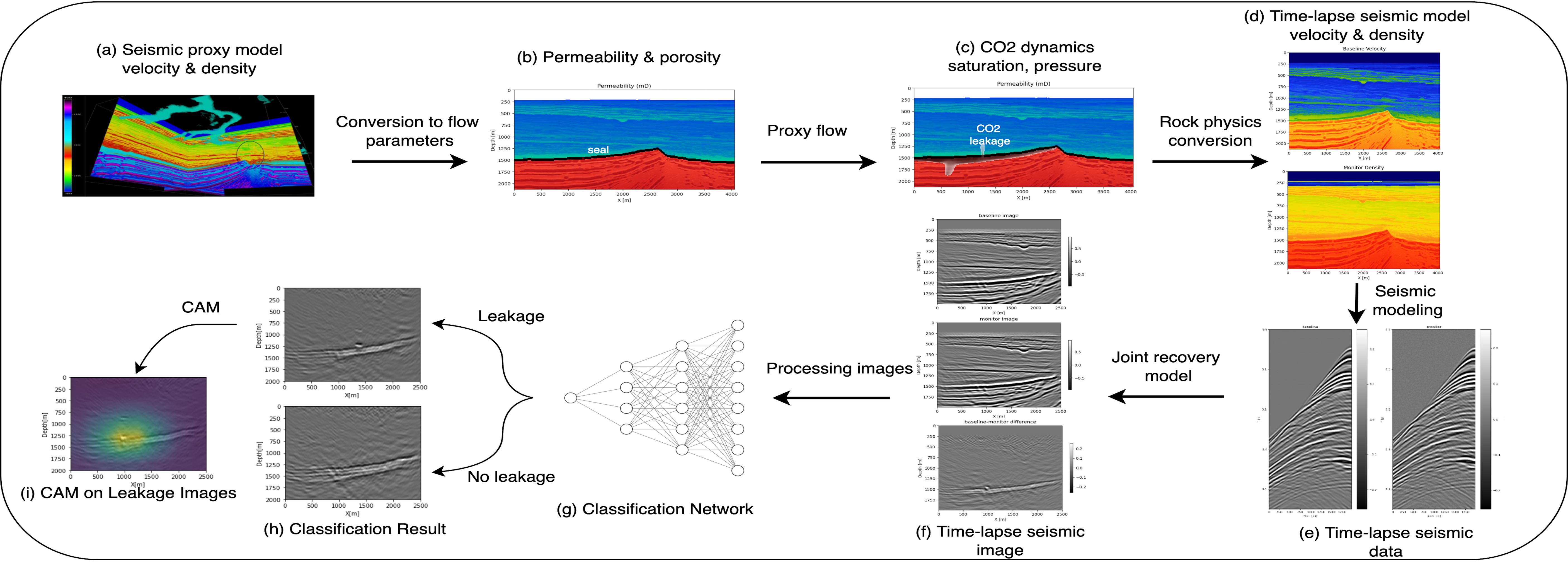}
\caption{\small{Workflow for CO${_2}$ Leakage Monitoring }}
\label{fig:figure1}
\end{figure*}

\section{Numerical Case Study}

To generate the training dataset of CO${_2}$ plume evolution, we used five 2D vertical slices extracted from the 3D Compass velocity model \cite{eage:/content/papers/10.3997/2214-4609.20148575} shown in Fig.~\ref{fig:figure1}(a). This model is a synthetic but realistic model representative of the complex geology of the southeast of the North Sea. The dimension of each model (slice) used in our work is $2131$ X $4062$ m$^2$. We used FwiFlow \cite{https://doi.org/10.1029/2019WR027032}, to simulate the CO${_2}$ flow dynamics and JUDI~\cite{witte2018alf} to model the seismic data and compute the seismic images of the subsurface.

\begin{table}[t]
\centering
\begin{tabular}{llllll}
\small{Hyperparameters}  & \small{VGG16} & \small{ResNet34} & \small{ViT}& \small{Swin}   \\ \Xhline{1.5pt}
\small{Batch Size} & \small{8}  & \small{8} & \small{8} & \small{8}  \\ 
\small{Learning Rate} & \small{$5$x$10^{-5}$} & \small{$6$x$10^{-3}$}  & \small{$4$x$10^{-3}$}  & \small{$10^{-3}$} \\ 
\small{Exp Decay Rate($\gamma$)}   & \small{$0.95$}   & \small{$0.92$}  & \small{$0.98$}  & \small{$0.98$}   \\ 
\end{tabular}
\caption{\small{Training hyperparameters for the four models. All models were trained with the same number of epochs and optimizer.}}
\label{table:hyperparameter table}
\end{table}

\subsection{Time-lapse reservoir and seismic simulation}

We consider a realistic two well setting with a fixed injection well injecting CO${_2}$ and a production well extracting brine from subsurface storage reservoir. Injection of supercritical CO${_2}$ into saline aquifers is an example of multi-phase flow in porous media. While we understand more complicated geothermal, geochemical and geomechanical process may eventually be considered to model the CO${_2}$ dynamics, we follow the two-phase immiscible incompressible flow physics, which in its leading order describes the process of supercritical CO${_2}$ displacing brine in the pore space of the rock. The system is governed by conservation of mass and Darcy's law. We refer to the existing literature \cite{https://doi.org/10.1029/2019WR027032, WEN2021103223} \cite{https://doi.org/10.1029/2019WR027032} for more details about this physical system.

Using empirical relation and the Kozeny-Carman equation\cite{https://doi.org/10.1029/2005GL025134}, the acoustic properties (velocity and density) from the Compass model were converted into permeability and porosity (Fig.~\ref{fig:figure1}(b)) to simulate the multi-phase flow (CO${_2}$ and brine in porous media) in the reservoir. We used FwiFlow.jl \cite{https://doi.org/10.1029/2019WR027032} to solve multi-phase flow equations based on the finite volume method. We simulated the CO${_2}$ flow for a duration varying between $7$ to $12$ years (Fig.~\ref{fig:figure1}(c)). The reservoir was initially filled with saline water and we injected compressed CO${_2}$ at the rate of $1$ MT/day into the reservoir for all simulations. In order to mimic CO${_2}$ leakage, we then created a fracture at a random location along the top seal of the reservoir when the pressure induced by the CO${_2}$ injection reaches a threshold of $15$ MPa. We then converted back these simulated CO${_2}$ saturation snapshots over time into wave properties with the patchy saturation model \cite{avseth2010quantitative} to obtain time-lapse subsurface models (Fig.~\ref{fig:figure1}(d)). We used this model because at higher pressure condition, local fluid flow slows down resulting in an acoustic velocity trend which follows patchy saturation \cite{Li2018ExperimentalSA}.

Based on these models, we then simulated the baseline seismic survey corresponding to the initial stage (before the injection of CO${_2}$) and the monitor seismic survey corresponding to the final stage at the end of the reservoir simulation (Fig.~\ref{fig:figure1}(e)). As mentioned in the introduction, it is very difficult to exactly replicate the baseline and monitor surveys. In order to mimic the realistic scenario in the field, the baseline and monitor datasets were simulated using different acquisition geometries (position of the measurements). Finally, we recovered the time-lapse seismic images using JRM \cite{oghenekohwo2017EAGEitl, wason2016GEOPctl, yin2021SEGcts} to alleviate potential noise and inaccuracies in the seismic images in the case of non-replicated time-lapse surveys. These recovered images along with the label (leakage/no-leakage) serve as the input to the classification network. We generated a total of $1000$ leakage and $870$ no-leakage scenarios, and computed the baseline, monitor and difference images with the JRM method in each case.

\begin{table*}[th]
\centering
\begin{tabular}{llllll}
\Xhline{1.5pt}
\textbf{Model}  & \textbf{Accuracy} & \textbf{Precision} & \textbf{Recall}& \textbf{F1} & \textbf{ROC-AUC}   \\ \Xhline{1.5pt}
VGG16  & 0.920$\pm$ 0.089  & 0.941$\pm$ 0.133   & 0.921$\pm$ 0.081 & 0.927$\pm$ 0.075 & 0.920$\pm$ 0.076 \\ 
ResNet34 & \textbf{0.948$\pm$ 0.020}  & \textbf{0.982$\pm$ 0.028}  & \textbf{0.928$\pm$ 0.044} & \textbf{0.948$\pm$ 0.040} & \textbf{0.967$\pm$ 0.019} \\ 
ViT   & 0.857$\pm$ 0.018  & 0.910$\pm$ 0.102   & 0.820$\pm$ 0.098 & 0.859$\pm$ 0.036 & 0.923$\pm$ 0.023  \\ 
Swin   & 0.836$\pm$ 0.036  & 0.881$\pm$ 0.108   & 0.818 $\pm$ 0.078 & 0.841$\pm$ 0.076 & 0.909$\pm$ 0.007 \\ \Xhline{1.5pt}
\end{tabular}
\caption{\small{Comparison of performance (for precision and recall, positives represent leakage whereas negatives are no leakage) on the test dataset for our four neural networks. The highest performance for each metric is highlighted in bold.}}
\label{table:Evaluation table}
\end{table*}

\begin{figure*}[t]
\centering
\includegraphics[width=\textwidth] {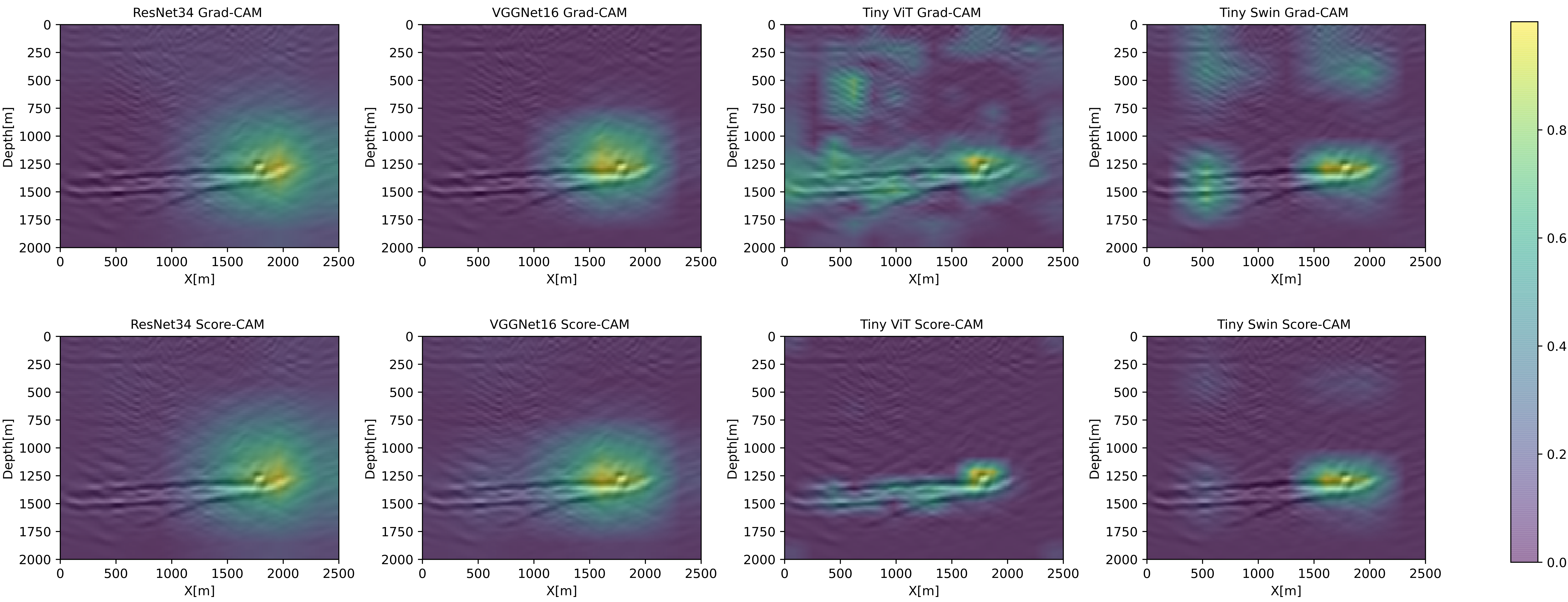}
\caption{\small{Grad-CAM and Score-CAM saliency maps overlayed on the corresponding input seismic image containing a CO${_2}$ plume from leakage. The CO${_2}$ plume can be seen on the seismic image as the high amplitude event at 1.3km depth and 1.8km in X.}}
\label{fig:figure2}
\end{figure*}

\subsection{Training}
The seismic difference images (difference between baseline and monitor recovery results) were converted to 224x224 gray-scale images with bi-linear interpolation and transformed into three channel images where each channel is a copy of the actual gray-scale image. For the classification, the image dataset was randomly split into an 80\% training set and 20\% test set. The training set was then further divided into two parts, one for model parameter updating (training) and another for hyperparameter tuning (validation). The training hyperparameters from this second part are summarized in Table~\ref{table:hyperparameter table}. For training, we replaced the last fully connected layers (classification layers) of each model with a new fully connected layer. We then trained the network (Fig.~\ref{fig:figure1}(g)) in two steps. First, we only trained the last classification layer, by freezing all the other layers, for 100 epochs. Since most of the layers are fixed and do not need gradient updates, this first stage is extremely cheap and computationally efficient. Second, we further trained the full model for an additional 30 epochs to allow fine-tuning of all layers for our specific classification task. Following standard practices in classification settings, we used a binary cross-entropy loss function and the Adam optimizer \cite{adam} for all models. Finally, after the training (Fig.~\ref{fig:figure1} (h)), we implemented the CAM based methods (Fig.~\ref{fig:figure1} (i)). We used the last convolutional layer activations for the CNN models, and the activations preceding the last attention layer for the transformer-based models.

\section{Analysis}

We show on Table~\ref{table:Evaluation table}, different performance metrics on our testing dataset, after training our four networks, with means and confidence intervals after 15 different runs. In detail, we show standard metrics such as accuracy, precision, and recall. Additionally, we also show F1 score~\cite{f1}, that combine recall and accuracy, and area under curve of receiver operating characteristic (ROC-AUC)~\cite{roc} to further evaluate the classification performance of models. We observe in Table~\ref{table:Evaluation table} that the CNN models outperform the transformer variants in all the metrics by a significant margin and that ResNet34 achieves the best performance in all the measures of evaluation. This result is consistent with the literature, hinting that despite being very accurate on a specific task, transformers do not generalize well with our modest sized dataset~\cite{Visiontransformer}. Additionally, we observe that all models lead to better precision compared to recall (more false negatives than false positives). This discrepancy can be attributed to the fact that certain leakage images have very small CO${_2}$ leakage areas (up to a single pixel) in the seismic images and are consequently very difficult to detect.

Second, we show in Fig.~\ref{fig:figure2} the CAM results of each model on a single seismic image from our test dataset. The high amplitude area shows the regions of the seismic images that are most important to the classifier. As expected, those heatmaps provide an explainable representation of the classification as the high amplitudes align with the CO${_2}$ leakage part of the seismic image. We observe that for the CNN, the saliency maps are well centered on the CO${_2}$ leakage portion despite being very coarse. Because of this coarseness, both Grad-CAM and score-CAM provide similar results. On the other hand, transformer-based networks lead to more focused saliency maps that target the location of the CO${_2}$ leakage extremely well. We observe in that case, the Score-CAM leads to reduction of aliases and noise compared to the Grad-CAM results. This can be linked to the potential presence of noise in the gradients of the transformers as the networks are very deep \cite{ScoreCAM}.

\section{Conclusion}

We have introduced an interpretable deep-learning method for CO${_2}$ leakage detection with very high accuracy on a synthetic but realistic model of a CO${_2}$ sequestration reservoir. First, we showed through four state-of-the-art models that we can detect potential CO${_2}$ leakage from the recovered time-lapse seismic images. Second, we demonstrated that CAM provides an interpretable and accurate visualization of the CO${_2}$ plume in case of leakage. Additionally, we showed that transformer-based models (ViT, Swin) led to more focused CAM and that Score-CAM provided cleaner and therefore more explainable heatmaps. On the other hand, we found that standard CNNs led to better classification results and therefore better leakage detection. In particular, ResNet model performed best and achieved a very high score above $90$\% in every evaluation metric. Future work will focus on improving the classification network to achieve higher accuracy in leakage detection and on refining the heatmaps for better explainability.

\section{Acknowledgments}

This research was carried out with the support of Georgia Research Alliance and partners of the ML4Seismic Center. The authors thank Philipp A. Witte at Microsoft for the constructive discussion.

\bibliography{aaai22.bbl}

\begin{thebibliography}{33}
\providecommand{\natexlab}[1]{#1}

\bibitem[{Arts et~al.(2008)Arts, Chadwick, Eiken, Thibeau, and
  Nooner}]{Arts2008TenYE}
Arts, R.~J.; Chadwick, A.; Eiken, O.; Thibeau, S.; and Nooner, S.~L. 2008.
\newblock Ten years' experience of monitoring CO2 injection in the Utsira Sand
  at Sleipner, offshore Norway.
\newblock \emph{First Break}, 26.

\bibitem[{Avseth, Mukerji, and Mavko(2010)}]{avseth2010quantitative}
Avseth, P.; Mukerji, T.; and Mavko, G. 2010.
\newblock \emph{Quantitative seismic interpretation: Applying rock physics
  tools to reduce interpretation risk}.
\newblock Cambridge university press.

\bibitem[{Ayeni and Biondi(2010)}]{Ayeni&Biondi}
Ayeni, G.; and Biondi, B. 2010.
\newblock Target-oriented joint least-squares migration/inversion of time-lapse
  seismic data sets.
\newblock \emph{Geophysics}, 75.

\bibitem[{Bradley(1997)}]{roc}
Bradley, A.~P. 1997.
\newblock The use of the area under the ROC curve in the evaluation of machine
  learning algorithms.
\newblock \emph{Pattern Recognition}, 30(7): 1145--1159.

\bibitem[{Chadwick et~al.(2010)Chadwick, Williams, Delepine, Clochard, Labat,
  Sturton, Buddensiek, Dillen, Nickel, Lima, Arts, Neele, and
  Rossi}]{doi:10.1190/1.3304820}
Chadwick, A.; Williams, G.; Delepine, N.; Clochard, V.; Labat, K.; Sturton, S.;
  Buddensiek, M.-L.; Dillen, M.; Nickel, M.; Lima, A.~L.; Arts, R.; Neele, F.;
  and Rossi, G. 2010.
\newblock Quantitative analysis of time-lapse seismic monitoring data at the
  Sleipner CO2 storage operation.
\newblock \emph{The Leading Edge}, 29(2): 170--177.

\bibitem[{Chinchor(1992)}]{f1}
Chinchor, N. 1992.
\newblock MUC-4 Evaluation Metrics.
\newblock In \emph{Proceedings of the 4th Conference on Message Understanding},
  MUC4 '92, 22–29. USA: Association for Computational Linguistics.
\newblock ISBN 1558602739.

\bibitem[{Costa(2006)}]{https://doi.org/10.1029/2005GL025134}
Costa, A. 2006.
\newblock Permeability-porosity relationship: A reexamination of the
  Kozeny-Carman equation based on a fractal pore-space geometry assumption.
\newblock \emph{Geophysical Research Letters}, 33(2).

\bibitem[{Dosovitskiy et~al.(2020)Dosovitskiy, Beyer, Kolesnikov, Weissenborn,
  Zhai, Unterthiner, Dehghani, Minderer, Heigold, Gelly, Uszkoreit, and
  Houlsby}]{Visiontransformer}
Dosovitskiy, A.; Beyer, L.; Kolesnikov, A.; Weissenborn, D.; Zhai, X.;
  Unterthiner, T.; Dehghani, M.; Minderer, M.; Heigold, G.; Gelly, S.;
  Uszkoreit, J.; and Houlsby, N. 2020.
\newblock An Image is Worth 16x16 Words: Transformers for Image Recognition at
  Scale.
\newblock \emph{arXiv}.

\bibitem[{E.~Jones et~al.(2012)E.~Jones, A.~Edgar, I.~Selvage, and
  Crook}]{eage:/content/papers/10.3997/2214-4609.20148575}
E.~Jones, C.; A.~Edgar, J.; I.~Selvage, J.; and Crook, H. 2012.
\newblock Building Complex Synthetic Models to Evaluate Acquisition Geometries
  and Velocity Inversion Technologies.
\newblock \emph{European Association of Geoscientists \& Engineers},
  cp-293-00580.

\bibitem[{Furre et~al.(2017)Furre, Eiken, Alnes, Vevatne, and
  Kiær}]{FURRE20173916}
Furre, A.-K.; Eiken, O.; Alnes, H.; Vevatne, J.~N.; and Kiær, A.~F. 2017.
\newblock 20 Years of Monitoring CO2-injection at Sleipner.
\newblock \emph{Energy Procedia}, 114: 3916--3926.
\newblock 13th International Conference on Greenhouse Gas Control Technologies,
  GHGT-13, 14-18 November 2016, Lausanne, Switzerland.

\bibitem[{He et~al.(2015)He, Zhang, Ren, and Sun}]{resnet}
He, K.; Zhang, X.; Ren, S.; and Sun, J. 2015.
\newblock Deep Residual Learning for Image Recognition.
\newblock \emph{arXiv}.

\bibitem[{IPCC(2018)}]{IPCC2018}
IPCC. 2018.
\newblock Global warming of 1.5°C. An IPCC Special Report on the impacts of
  global warming of 1.5°C above pre-industrial levels and related global
  greenhouse gas emission pathways, in the context of strengthening the global
  response to the threat of climate change, sustainable development, and
  efforts to eradicate poverty.
\newblock \emph{In Press}.

\bibitem[{Kingma and Ba(2015)}]{adam}
Kingma, D.~P.; and Ba, J. 2015.
\newblock Adam: A Method for Stochastic Optimization.
\newblock In \emph{ICLR (Poster)}.

\bibitem[{Li et~al.(2018)Li, Zhou, Li, Duguid, Que, Xue, and Tan}]{LI2018276}
Li, B.; Zhou, F.; Li, H.; Duguid, A.; Que, L.; Xue, Y.; and Tan, Y. 2018.
\newblock Prediction of CO2 leakage risk for wells in carbon sequestration
  fields with an optimal artificial neural network.
\newblock \emph{International Journal of Greenhouse Gas Control}, 68: 276--286.

\bibitem[{Li et~al.(2020)Li, Xu, Harris, and
  Darve}]{https://doi.org/10.1029/2019WR027032}
Li, D.; Xu, K.; Harris, J.~M.; and Darve, E. 2020.
\newblock Coupled Time-Lapse Full-Waveform Inversion for Subsurface Flow
  Problems Using Intrusive Automatic Differentiation.
\newblock \emph{Water Resources Research}, 56(8): e2019WR027032.
\newblock E2019WR027032 10.1029/2019WR027032.

\bibitem[{Liu et~al.(2021)Liu, Lin, Cao, Hu, Wei, Zhang, Lin, and
  Guo}]{Swintransformer}
Liu, Z.; Lin, Y.; Cao, Y.; Hu, H.; Wei, Y.; Zhang, Z.; Lin, S.; and Guo, B.
  2021.
\newblock Swin Transformer: Hierarchical Vision Transformer using Shifted
  Windows.
\newblock \emph{arXiv}.

\bibitem[{Lumley(2001)}]{doi:10.1190/1.1444921}
Lumley, D.~E. 2001.
\newblock Time-lapse seismic reservoir monitoring.
\newblock \emph{GEOPHYSICS}, 66(1): 50--53.

\bibitem[{Oghenekohwo and Herrmann(2017{\natexlab{a}})}]{oghenekohwo2017hrt}
Oghenekohwo, F.; and Herrmann, F.~J. 2017{\natexlab{a}}.
\newblock Highly repeatable time-lapse seismic with distributed Compressive
  Sensing{\textendash}-mitigating effects of calibration errors.
\newblock \emph{The Leading Edge}, 36(8): 688--694.
\newblock (The Leading Edge).

\bibitem[{Oghenekohwo and
  Herrmann(2017{\natexlab{b}})}]{oghenekohwo2017EAGEitl}
Oghenekohwo, F.; and Herrmann, F.~J. 2017{\natexlab{b}}.
\newblock Improved time-lapse data repeatability with randomized sampling and
  distributed compressive sensing.
\newblock In \emph{EAGE Annual Conference Proceedings}.
\newblock (EAGE, Paris).

\bibitem[{Ringrose(2020)}]{ringrose2020store}
Ringrose, P. 2020.
\newblock \emph{How to store CO2 underground: Insights from early-mover CCS
  Projects}, volume 129.
\newblock Springer.

\bibitem[{Ringrose et~al.(2013)Ringrose, Mathieson, Wright, Selama, Hansen,
  Bissell, Saoula, and Midgley}]{RINGROSE20136226}
Ringrose, P.; Mathieson, A.; Wright, I.; Selama, F.; Hansen, O.; Bissell, R.;
  Saoula, N.; and Midgley, J. 2013.
\newblock The In Salah CO2 Storage Project: Lessons Learned and Knowledge
  Transfer.
\newblock \emph{Energy Procedia}, 37: 6226--6236.
\newblock GHGT-11 Proceedings of the 11th International Conference on
  Greenhouse Gas Control Technologies, 18-22 November 2012, Kyoto, Japan.

\bibitem[{Russakovsky et~al.(2014)Russakovsky, Deng, Su, Krause, Satheesh, Ma,
  Huang, Karpathy, Khosla, Bernstein, Berg, and Fei-Fei}]{imagenet}
Russakovsky, O.; Deng, J.; Su, H.; Krause, J.; Satheesh, S.; Ma, S.; Huang, Z.;
  Karpathy, A.; Khosla, A.; Bernstein, M.; Berg, A.~C.; and Fei-Fei, L. 2014.
\newblock ImageNet Large Scale Visual Recognition Challenge.

\bibitem[{Selvaraju et~al.(2019)Selvaraju, Cogswell, Das, Vedantam, Parikh, and
  Batra}]{Grad-CAM}
Selvaraju, R.~R.; Cogswell, M.; Das, A.; Vedantam, R.; Parikh, D.; and Batra,
  D. 2019.
\newblock Grad-{CAM}: Visual Explanations from Deep Networks via Gradient-Based
  Localization.
\newblock \emph{International Journal of Computer Vision}, 128(2): 336--359.

\bibitem[{Simonyan and Zisserman(2014)}]{vgg}
Simonyan, K.; and Zisserman, A. 2014.
\newblock Very Deep Convolutional Networks for Large-Scale Image Recognition.
\newblock \emph{arXiv}.

\bibitem[{Vaswani et~al.(2017)Vaswani, Shazeer, Parmar, Uszkoreit, Jones,
  Gomez, Kaiser, and Polosukhin}]{transformer}
Vaswani, A.; Shazeer, N.; Parmar, N.; Uszkoreit, J.; Jones, L.; Gomez, A.~N.;
  Kaiser, L.; and Polosukhin, I. 2017.
\newblock Attention Is All You Need.
\newblock \emph{arXiv}.

\bibitem[{Wang et~al.(2019)Wang, Wang, Du, Yang, Zhang, Ding, Mardziel, and
  Hu}]{ScoreCAM}
Wang, H.; Wang, Z.; Du, M.; Yang, F.; Zhang, Z.; Ding, S.; Mardziel, P.; and
  Hu, X. 2019.
\newblock Score-CAM: Score-Weighted Visual Explanations for Convolutional
  Neural Networks.
\newblock \emph{arXiv}.

\bibitem[{Wason, Oghenekohwo, and Herrmann(2017)}]{wason2016GEOPctl}
Wason, H.; Oghenekohwo, F.; and Herrmann, F.~J. 2017.
\newblock Low-cost time-lapse seismic with distributed compressive
  sensing{\textendash}-Part 2: impact on repeatability.
\newblock \emph{Geophysics}, 82(3): P15--P30.
\newblock (Geophysics).

\bibitem[{Wen, Tang, and Benson(2021)}]{WEN2021103223}
Wen, G.; Tang, M.; and Benson, S.~M. 2021.
\newblock Towards a predictor for CO2 plume migration using deep neural
  networks.
\newblock \emph{International Journal of Greenhouse Gas Control}, 105: 103223.

\bibitem[{Witte et~al.(2019a)Witte, Louboutin, Kukreja, Luporini, Lange,
  Gorman, and Herrmann}]{witte2018alf}
Witte, P.~A.; Louboutin, M.; Kukreja, N.; Luporini, F.; Lange, M.; Gorman,
  G.~J.; and Herrmann, F.~J. 2019a.
\newblock A large-scale framework for symbolic implementations of seismic
  inversion algorithms in Julia.
\newblock \emph{Geophysics}, 84(3): F57--F71.
\newblock (Geophysics).

\bibitem[{Yin, Louboutin, and Herrmann(2021)}]{yin2021SEGcts}
Yin, Z.; Louboutin, M.; and Herrmann, F.~J. 2021.
\newblock Compressive time-lapse seismic monitoring of carbon storage and
  sequestration with the joint recovery model.
\newblock In \emph{SEG Technical Program Expanded Abstracts}, 3434--3438.
\newblock (IMAGE, Denver).

\bibitem[{Yosinski et~al.(2014)Yosinski, Clune, Bengio, and
  Lipson}]{transfer_learning}
Yosinski, J.; Clune, J.; Bengio, Y.; and Lipson, H. 2014.
\newblock How transferable are features in deep neural networks?
\newblock \emph{arXiv}.

\bibitem[{Zhou et~al.(2015)Zhou, Khosla, Lapedriza, Oliva, and Torralba}]{cam}
Zhou, B.; Khosla, A.; Lapedriza, A.; Oliva, A.; and Torralba, A. 2015.
\newblock Learning Deep Features for Discriminative Localization.
\newblock \emph{arXiv}.

\bibitem[{Zhou et~al.(2019)Zhou, Lin, Zhang, Wu, Wang, Dilmore, and
  Guthrie}]{ZHOU2019102790}
Zhou, Z.; Lin, Y.; Zhang, Z.; Wu, Y.; Wang, Z.; Dilmore, R.; and Guthrie, G.
  2019.
\newblock A data-driven CO2 leakage detection using seismic data and
  spatial-temporal densely connected convolutional neural networks.
\newblock \emph{International Journal of Greenhouse Gas Control}, 90: 102790.

\end{thebibliography}
\end{document}